\documentclass[twocolumn,secnumarabic,showpacs,showkeys,amsmath,amssymb,floatfix]{revtex4-1}

\usepackage[english]{babel}
\usepackage{amsfonts}
\usepackage{graphicx}
\usepackage{color}
\usepackage{bm}
\input colordvi

\newcommand{\fmf}{\affiliation{Faculty of Mathematics and Physics, University of Ljubljana, SI-1000 Ljubljana, Slovenia}}
\newcommand{\ijs}{\affiliation{J. Stefan Institute, SI-1000 Ljubljana, Slovenia}}
\newcommand{\lmu}{\affiliation{Department of Physics and Arnold Sommerfeld Center for Theoretical Physics, Ludwig-Maximilians-Universit\"at M\"unchen, D-80333 M\"unchen, Germany}}

\begin{document}

\title{Two holes in the $t$--$J$ model form a bound state for any nonzero $J/t$}

\author{L. \surname{Vidmar}}
\lmu
\ijs

\author{J. \surname{Bon\v ca}}
\fmf
\ijs

%
%
%

\date{\today}
\begin{abstract}
Determination of the parameter regime in which two holes in the $t$--$J$ model form a bound state represents a long standing open problem in the field of strongly correlated systems.
By applying and systematically improving the exact diagonalization method defined over a limited functional space (EDLFS), we show that the average distance between two holes scales as $\langle d \rangle \sim 2 (J/t)^{-1/4}$ for $J/t < 0.15$, therefore providing strong evidence that two holes in the $t$--$J$ model form the bound state for any nonzero $J/t$.
However, the symmetry of such bound pair in the ground state is $p$--wave.
This state is consistent with phase separation at finite hole filling, as observed in a recent study [Maska {\it et al}, Phys. Rev. B {\bf 85}, 245113 (2012)].
\end{abstract}

\keywords{Strongly correlated systems, bound states, phase separation}
\pacs{71.27.+a, 71.38.Mx, 74.20.Rp}

\maketitle

\section{Introduction}

In the context of strongly correlated models, one of the fundamental problems emerging soon after the discovery of high--${\rm T_c}$ cuprates was to explore whether two holes doped into the ${\rm CuO_2}$ plane form a bound state~\cite{bonca89}.
Despite the intensive work on this problem in the last 20 years, investigation of binding even within the single band models, like the $t$--$J$ model, continued to represent a challenging task~\cite{*[{See, e.g., }] [{ and references therein.}] dagotto_rev}.
After extensive analytical as well as numerical efforts devoted to the two--hole problem in the $t$--$J$ model~\cite{dagotto_rev,chernyshev98,barentzen99,riera98,wrobel98,leung02,tohyama04}, a general consensus has been established that the bound pair with a $d$--wave symmetry represents the ground state in a parameter range $J/t \gtrsim 0.4$.
Nevertheless, in the regime $J/t \ll 1$ the symmetry of the ground state crosses to $p$--wave, raising doubts of many authors about the relevance of the $t$--$J$ model for high--${\rm T_c}$ superconductivity.
Moreover, the average distance between holes increases with decreasing $J/t$, which impedes accurate quantum mechanical treatments of the problem.
Since recently, there has been no consensus on the parameter regime in which two holes in the $t$--$J$ model form a bound state.
In particular, detection of the bound state in the regime $J/t \ll 1$ where the possible $p$--wave bound pair may emerge
has remained an open problem.

Due to a steady development of modern numerical techniques to study strongly correlated systems in two dimensions, the physics of the $t$--$J$ model has recently experienced a revival since the nature of its ground state at finite doping is still a controversial issue.
The main question in this context is whether the ground state is homogeneous with the possible signatures of $d$--wave pairing~\cite{sorella02,hu12} or it possesses inhomogeneous state, e.g., phase separation~\cite{maska12} or stripe order~\cite{corboz11}.
So far, no agreement on this issue has been reached yet.
In this respect, accurate solutions of few--hole systems may provide a valuable information about the weak--doping limit of the model.
In particular, we show that properties of the bound state in the $J/t \ll 1$ limit are consistent with the phase--separated state of hole--rich ferromagnetic and hole--depleted antiferromagnetic regions, observed recently at finite doping~\cite{maska12}.

We consider the $t$--$J$ model with two holes on the square lattice
\begin{equation}
H = -t \sum_{\langle {\bf ij}\rangle,s}(\tilde c^\dagger_{{\bf i},s} \tilde c_{{\bf j},s} +\mathrm{H.c.}) +
\sum_{\langle {\bf ij}\rangle } J ( {\bf S}_{\bf i} {\bf S}_{\bf j} - \frac{1}{4}\tilde{n}_{\bf i} \tilde{n}_{\bf j} )
\label{ham}
\end{equation}
where $\tilde c_{{\bf i},s} = c_{{\bf i},s}(1 - n_{{\bf i},-s})$ is a projected fermion operator, $t$ represents nearest neighbor overlap integral, the sum $\langle \bf ij \rangle$  runs over pairs of nearest neighbors and $\tilde n_{\bf i} = n_{\bf{i},\uparrow} + n_{\bf{i},\downarrow} - 2 n_{\bf{i},\uparrow} n_{\bf{i},\downarrow}$ is a projected electron number operator.
We set $t=1$ throughout the work.
By solving to the two--hole problem in the whole regime of $J < 1$, we benchmark the EDLFS method as an extremely efficient technique to study the weak--doping limit of the $t$--$J$ model.

\section{EDLFS method}

We apply the exact diagonalization defined method over a limited functional space (EDLFS) for the $t$--$J$ model~\cite{bonca2,vidmar09b}.
The construction of the functional space starts from a N\' eel state with two holes located on neighboring ${\rm Cu}$ sites~\cite{vidmar09b}, which represents a parent state of a translationally invariant state with ${\bf k} = (0, 0)$
\begin{equation}
\vert \phi^{(0)}{\rangle}_{\alpha} = \sum_{\boldsymbol{\beta}}(-1)^{M_{\alpha}(\boldsymbol{\beta})}
c_{0,\sigma}c_{\boldsymbol{\beta},-\sigma}\vert \mbox{N\' eel}\rangle, \label{parent}
\end{equation}
where sum over $\boldsymbol{\beta}$ runs over four nearest neighbors in the case of $d$--wave symmetry and over two in the case of $p_{x(y)}$--wave symmetry.
The parameter $M_{\alpha} (\boldsymbol{\beta})$, $\alpha \in \{ d,p \}$ sets the appropriate sign of the wavefunctions.
Four wavefunctions contributing to the sum over $\boldsymbol{\beta}$ in Eq.~(\ref{parent}) as well as the definition of $M_{\alpha} (\boldsymbol{\beta})$ are shown in Fig.~1 of Ref.~\cite{vidmar09b}.
We generate new parent states by applying the generator of states
\begin{equation}
\left\{|\phi_{j}^{(n_h)} \rangle \right\} = \left[ H_{kin} + H_J^{\mbox{\tiny off}} \right]^{n_h} \vert \phi^{(0)}{\rangle}_{\alpha} \label{basis}
\end{equation}
where $n_h=0,...,N_h$ and $H_{kin}$, $H_J^{\mbox{\tiny off}}$ represents off--diagonal parts of Eq.~(\ref{ham}).
Full Hamiltonian is diagonalized within the limited functional space taking explicitly into account translational symmetry.

The method has been successfully applied to calculation of the ground state of the $t$--$J$ model with one~\cite{bonca2,vidmar11b} and two doped holes~\cite{vidmar09b,maska12,bonca12}, as well as extended to studies of the $t$--$J$ model with lattice degrees of freedom~\cite{bonca3,vidmar09b,vidmar11c}.
One of the significant advantages of the method represents its ability to study large hole distances up to $N_h+1$.
We take advantage of this property in Sec.~\ref{sec_c} where we detect the emergence of the bound state.
A systematic finite size scaling of the results in the extreme $J \ll 1$ limit will be presented in Sec.~\ref{sec_d}.

\section{Detection of the bound state} \label{sec_c}

\begin{figure}[!tbh]
\includegraphics[width=0.9\columnwidth,clip]{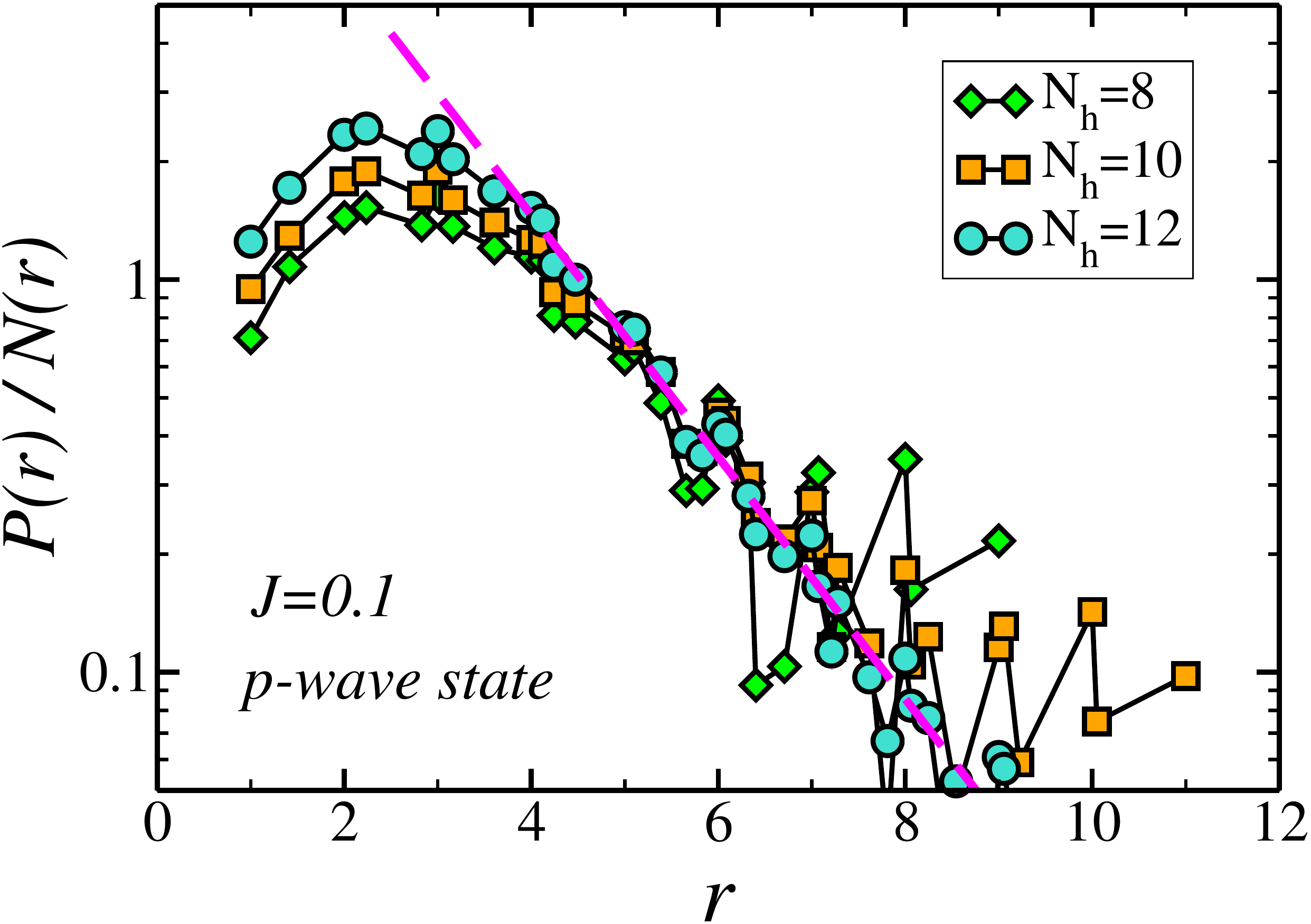}
\caption{(Color online)
$P(r)/N(r)$ in the ground state the $t$--$J$ model with $p$--wave symmetry at $J=0.1$.
Dashed line represent a fit to the data at $r\geq 4$ for $N_h=12$ according to Eq.~(\ref{expdecay}), which gives $\xi=1.41$.
}
\label{fig1}
\end{figure}

We first focus on the problem how to detect the emergence of the bound state at small $J$.
We define the hole--hole probability function in the ground state as
\begin{equation}
P(r) = \langle \psi_{\rm GS} \vert \sum_{\langle \bf i\not = j \rangle } n^h_{\bf i}n^h_{\bf j} \delta\left [ \vert {\bf i-j}\vert -r\right ] \vert \psi_{\rm GS}  \rangle, \label{Pr}
\end{equation}
where $n^h_{\bf i}$ represents the hole number operator.
If the holes form the bound state, we expect that $P(r)$ exhibits an exponential decay at large $r$.
However, such condition for the bound state is not necessarily enough since we also have to prove that the functional space generator defined in Eq.~(\ref{basis}) does not systematically favor states at smaller hole distances.
For this purpose, we define the distribution function $N(r)$
\begin{equation}
N(r) = \langle \tilde{\psi} \vert \sum_{\langle \bf i\not = j \rangle } n^h_{\bf i}n^h_{\bf j} \delta\left [ \vert {\bf i-j}\vert -r\right ] \vert \tilde{\psi} \rangle, \label{Nr}
\end{equation}
which calculates the probability for two holes to be at a distance $r$ provided that all states within our functional space are occupied with the equal probability.
Therefore, the strict condition for the existence of the bound state within the EDLFS method can be expressed as
\begin{equation}
\frac{P(r)}{N(r)} \sim e^{-r/\xi} \label{expdecay}
\end{equation}
when $r \gg \xi$.
Similar arguments have been recently applied in a two--hole study of a three--band model~\cite{lau11b}.
We plot $P(r)/N(r)$ at $J=0.1$ in Fig.~\ref{fig1}, which clearly reveals the existence of the bound state when $N_h$ is increased.
Remarkably, the figure reveals an exponential decay of the hole--hole probability at large distances with $\xi\sim1.4$.
The exponential decay can be efficiently detected using the EDLFS method where the maximal distance between two holes can be as large as $N_h+1=13$.
In contrary, the investiagtion of two--hole problems by means of exact diagonalization is restricted to bound pairs with a radius of only few lattice distances.
On a $N$--site cluster, the largest possible distance between two holes in the full Hilbert space is $\langle d \rangle_{\rm max}=\sqrt{N/2}$, leading to $\langle d \rangle_{\rm max}=4$ for $N=32$~\cite{leung02}.
As we shall show in the following, $\langle d \rangle \gtrsim 4$ in the limit $J \ll 1$ hence states with considerably larger inter--hole distances need to be taken into account.

Until now, there has been no consensus about the possible emergence of the bound state at $J=0.1$ since the majority of previous studies suggested that the state of two holes at $J=0.1$ is unbound~\cite{poilblanc93,boninsegni93,riera98,wrobel98,leung02}.
Nevertheless, we make a step further and show that two holes form the bound state for {\it any} finite $J$.
For this purpose, we need to calculate the average distance between holes.

\section{Average distance between holes} \label{sec_d}

\begin{figure}[!tbh]
\includegraphics[width=0.9\columnwidth,clip]{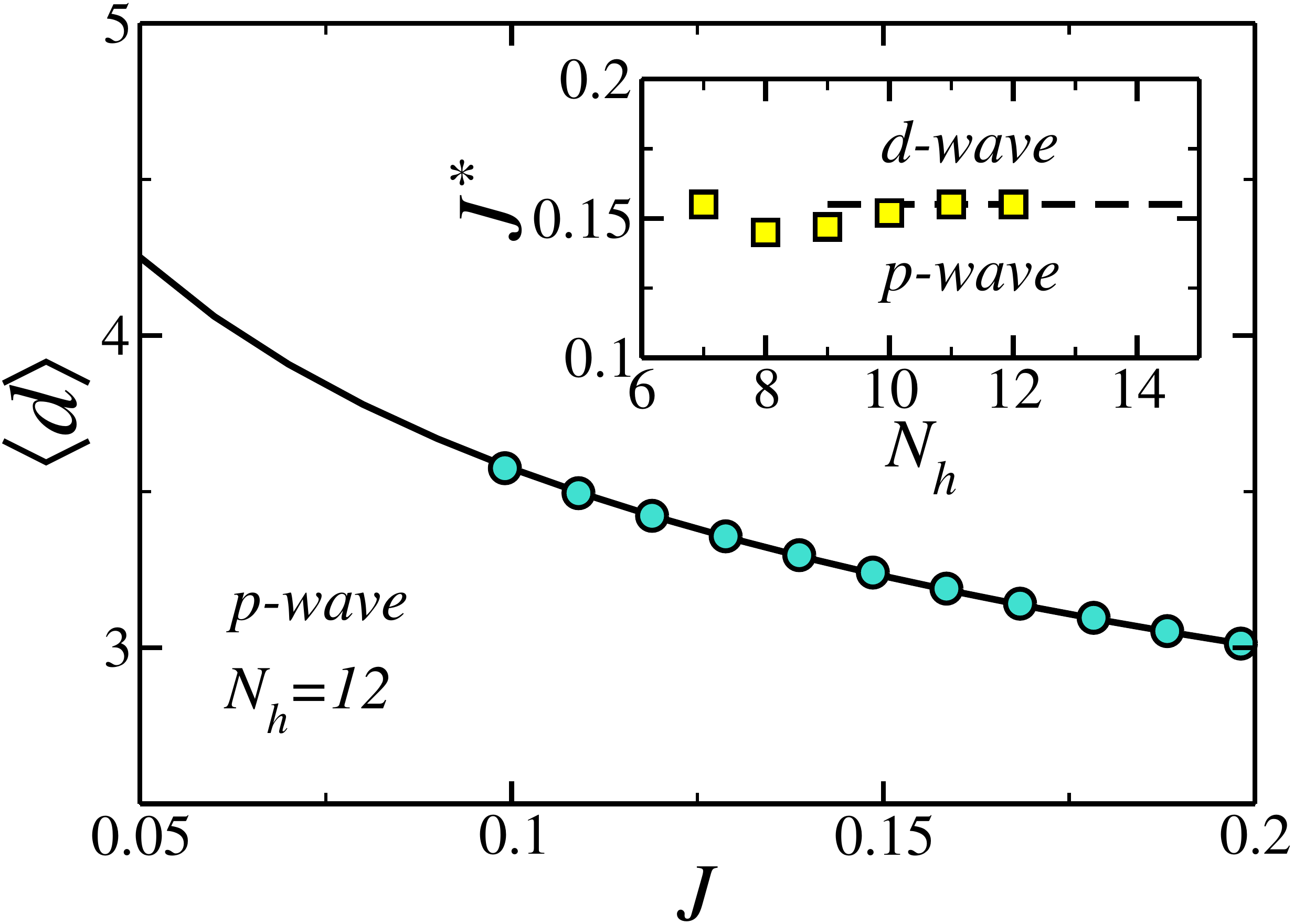}
\caption{(Color online)
Main panel: 
$\langle d \rangle$ vs $J$ for the $p$--wave bound pair.
Circles represent numerical results using the EDLFS method while the solid line represents a fit to the data $\langle d \rangle = \alpha J^{-1/4}$ with $\alpha=2.01$.
Inset: crossover $J^*$ between the $p$-- and $d$--wave ground state vs $N_h$.
}
\label{fig2}
\end{figure}

The average distance between holes is defined as
\begin{equation}
\langle d \rangle = \sum_r r P(r),
\end{equation}
where $P(r)$ has been introduced in Eq.~(\ref{Pr}).
In Fig.~\ref{fig2}, we show a scaling at $J_0 < J \ll 1$
\begin{equation}
\langle d \rangle = \alpha J^{-1/4}, \label{dscaling}
\end{equation}
where $\alpha \sim 2$ and $J_0$ represents the minimal $J$ for which our results are not biased due to finite--size effects.
The ansatz of the scaling in Eq.~(\ref{dscaling}) is motivated from the single--hole studies at $J \ll 1$, where it was shown that the ferromagnetic radius of the Nagaoka polaron scales as $J^{-1/4}$~\cite{white01}.
If two holes behave according to the same scaling, we may expect that they reside within the same ferromagnetic bubble.
We discuss this issue in more detail in Sec.~\ref{sec_conc}.
At this point, assuming that we have found the scaling of $\langle d \rangle$ in the limit $J \ll 1$, we need to show that it is robust against finite--size effects of the numerical method.

\begin{figure}[!tbh]
\includegraphics[width=0.9\columnwidth,clip]{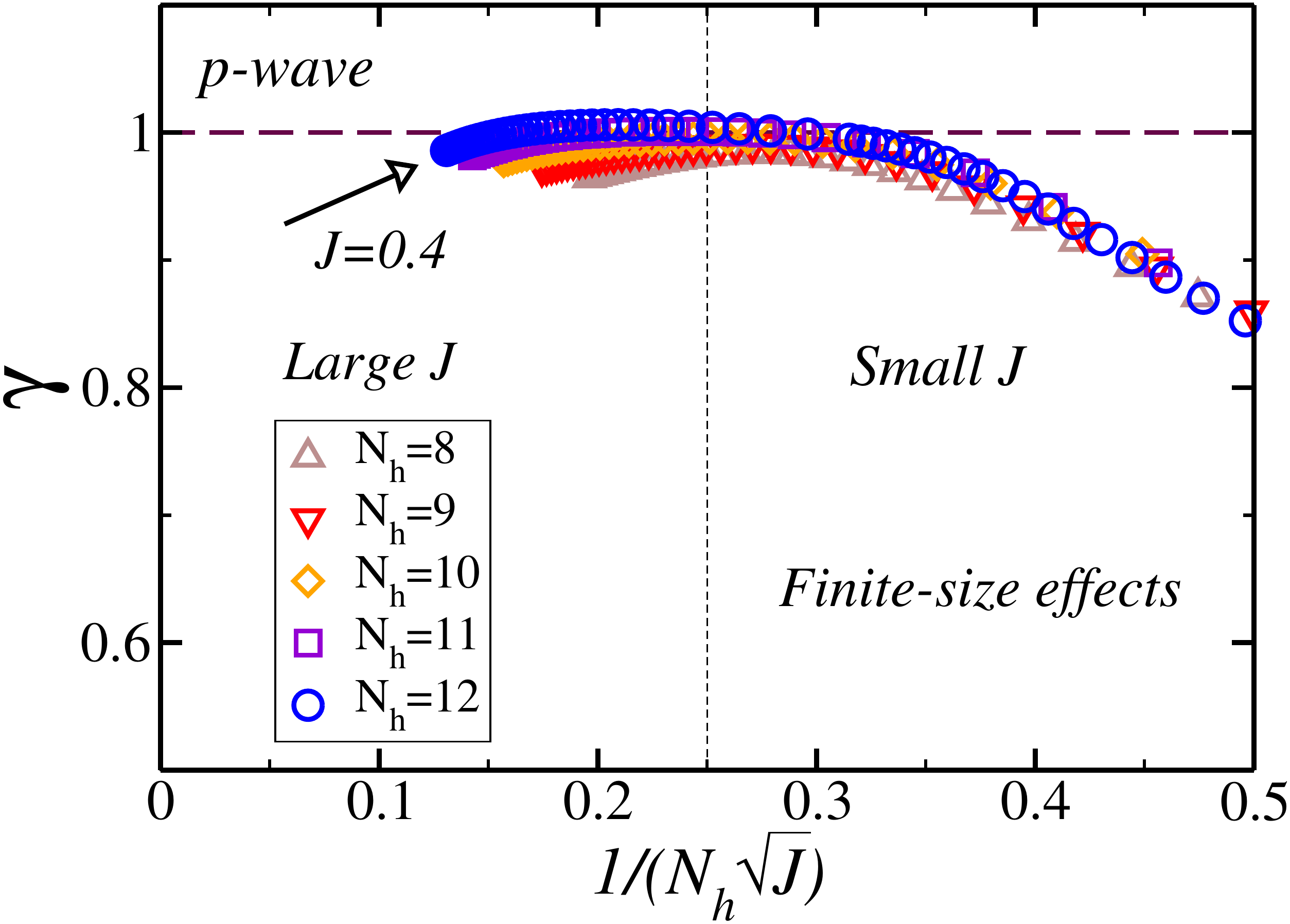}
\caption{(Color online)
$\gamma = \langle d \rangle / 2J^{-1/4}$ vs $1/(N_h\sqrt{J})$ for the $p$--wave state at $J<0.4$.
Note that the $p$--wave state represents the ground state of the $t$--$J$ model for $J \lesssim 0.15$, see also Fig.~\ref{fig2}.
}
\label{fig3}
\end{figure}

For this purpose we focus on the construction of the EDLFS method to detect the region of parameters for which our results may be considerably influenced by finite--size effects.
In the generation of the functional space, Eq.~(\ref{basis}), the holes propagate in each step of the generation in all possible directions, resembling the process of random walk where the time unit is determined by the parameter $N_h$.
We define $\langle r_0 \rangle$ as the average distance between two holes when all the states within the given functional space are occupied with the same probability, see also Eq.~(\ref{Nr}).
According to the random walk argument, we expect that $\langle r_0 \rangle$ scales with $N_h$ as
\begin{equation}
\langle r_0^2 \rangle  \sim N_h. \label{r0}
\end{equation}
In context of calculation of the average distance between the holes in the ground state $\langle d \rangle$, we should limit the calculation to distances lower than $\langle r_0 \rangle$.
Eq.~(\ref{r0}) suggests that the upper boundary of $\langle d \rangle$ should be determined as
\begin{equation}
\langle d \rangle <  \langle r_0 \rangle \sim \sqrt{N_h}. \label{dNh}
\end{equation}
%
By applying the scaling of $\langle d \rangle$ as given in Eq.~(\ref{dscaling}), we may rewrite Eq.~(\ref{dNh}) as
\begin{equation}
2 J^{-1/4}  <  \sqrt{N_h}, \label{jnh}
\end{equation}
which leads to
\begin{equation}
\frac{1}{N_h \sqrt{J}} <  \frac{1}{4}. \label{constraint}
\end{equation}
Therefore, Eq.~(\ref{constraint}) provides the constraint for our numerical calculation, i.e., it estimates for a given $N_h$ the lower bound of $J$ down to which we may expect converged results.

In Fig.~\ref{fig3} we show $\gamma = \langle d \rangle / 2J^{-1/4}$ vs $1/(N_h\sqrt{J})$.
If $1/(N_h\sqrt{J}) < 0.25$, we expect $\gamma \to 1$ if the scaling of Eq.~(\ref{dscaling}) is correct.
By increasing $N_h$ from 8 to 12, we indeed observe such result.
Note that increasing $N_h$ implies the decrease of $J_0$.
According to Eq.~(\ref{jnh}), $J_0 = 16/N_h^2$ represents the lower bound of the regime where finite--size effects are not expected to influence the results.
Keeping $\gamma \approx 1$ during the process of increasing $N_h$ (decreasing $J_0$) suggests that the scaling of Eq.~(\ref{dscaling}) is valid for any small and finite $J$.
Nevertheless, when $J < J_0$ we observe strong deviations from this scaling, as shown on the right side of Fig.~\ref{fig3}.




\section{Discussion and Outlook} \label{sec_conc}

Results presented in Fig.~\ref{fig3} provide s strong evidence that two holes in the $t$--$J$ model form a bound state for any finite $J$.
A natural question arising from this study concerns the possible extension of the two--hole problem to finite doping regime.
Results based on the recently proposed Ising version of the $t$--$J$ model~\cite{maska09} (which gives comparable results in the isotropic $t$--$J$ model in the limit $J \ll 1$), reveal a nearly same scaling of $\langle d \rangle$ vs $J$~\cite{maska12}.
In this picture, a so--called Nagaoka bipolaron is formed where the doped holes reside in a single bubble with a ferromagnetic (FM) spin alignment, while the rest of the system represents a hole--depleted antiferromagnetic region.
When the hole doping is further increased, holes keep residing within the same FM bubble, therefore leading to a phase separation at finite doping~\cite{maska12}.
Such phase separated state at $J \ll 1$ is driven by a minimization of hole kinetic energy, which leads to a ferromagnetically polarized cloud (bubble) of surrounding spins.
When for a fixed $J$ the hole doping is increased, the size of the FM bubble increases at the expense of the reduced hole--depleted AFM region, unless the whole spin sector is fully polarized~\cite{maska12}.


We now turn back to the two--holes studies and focus on the physically relevant regime of the $t$--$J$ model at $J\sim 0.3-0.4$.
When $J$ increases towards $0.3$, there is a competition in the ground state of the two--hole $t$--$J$ model between the $p$--wave state studied in this work and the $d$--wave state.
Our results strongly suggest that this $p$--wave bound state will unlikely lead to superconductivity at finite doping.
According to the EDLFS method, the symmetry of the bound pair in the ground state changes from $p$--wave to $d$--wave at $J^*\sim 0.15$ (see also the inset of Fig.~\ref{fig2}).
Note that this value of $J^*$ is notably lower than predicted from other studies~\cite{boninsegni93,prelovsek93,poilblanc94b,riera98,leung02}.
However, when adding realistic next--nearest--neighbor hopping terms to the $t$--$J$ model, $J^*$ increases~\cite{leung02,martins01} and possibly exceeds the regime $J\sim 0.3-0.4$.
This posses some serious challenges to the applicability of the extended $t$--$J$ models to describe high--${\rm T_c}$ superconductivity.

There are at least two directions recently investigated which may overcome these problems.
The first one concerns more general models of the ${\rm CuO_2}$ plane beyond the single band $t$--$J$ model.
Lately, a detailed exact numerical study of the two--hole problem within a projected three--band model was carried out~\cite{lau11b}.
However, no clear signatures of binding have been found on a 32 site cluster.
Therefore, the issue of binding of doped holes in strongly correlated multi--band models remains an open problem.
Another direction of investigation is to add lattice degrees of freedom to strongly correlated systems~\cite{wellein96,sakai97,hague07,huang11,miranda11,alexandrov12,vidmar09b}.
In this context, it has been shown that a coupling to transverse polarization of lattice vibrations stabilizes the $d$--wave symmetry of the bound state~\cite{vidmar09b}.
The major obstacle in studying strongly correlated systems with electron--phonon interaction in nonperturbative regime, is that there exist only a few reliable methods to treat such complex systems.
A recent study~\cite{vidmar10a} has nevertheless indicated that some very interesting physical properties of the bound pair may emerge at intermediate values of e--ph coupling, i.e., in the regime between the weak and strong electron--phonon interaction.
Since this regime of parameters represent a widely unexplored field, we believe it would be worth focusing on it in more detail in the future.

\begin{acknowledgments}
We thank M. Mierzejewski, M. M. Ma\'{s}ka, T. Tohyama and O. P. Sushkov for fruitful discussions.
We acknowledge support by the P1-0044 of ARRS, Slovenia.
J.B expresses gratitude for the support of CINT user program, Los Alamos National Laboratory, NM USA and Gordon Godfrey bequest of UNSW, Sydney Australia where part of this work has been performed.
\end{acknowledgments}

\bibliography{../../../../references}

\end{document}